\documentclass[aps,prd,final,twocolumn,letterpaper,nofootinbib]{revtex4-1}

\usepackage{ylmath}
\usepackage{qmech}

\usepackage{graphicx}
\usepackage{tikz}

\usepackage{color}

\usepackage[colorlinks]{hyperref}

\newcommand{\ylnotes}[1]{}

\begin{document}

\title{Scheme for generalized maximally localized Wannier functions in one dimension}
\author{Yuri D. Lensky}
\email{ydl@mit.edu}
\affiliation{Physics Department, Massachusetts Institute of Technology, 77 Massachusetts Avenue, Cambridge MA 02139}
\author{Colin Kennedy}
\affiliation{Center for Ultracold Atoms, Massachusetts Institute of Technology, 77 Massachusetts Avenue, Cambridge MA 02139}
\date{\today}

\begin{abstract}
Maximally localized Wannier functions are the key tool for a variety of physical applications of Bloch states. Here we develop a simple and exact procedure to construct maximally localized Wannier functions for one dimensional periodic potentials of arbitrary form. As opposed to relatively complex numerical minimization approaches that may return somewhat different results depending on implementation and running conditions, this computationally straightforward method guarantees a unique (and optimal) result on each run. These features make it a useful vehicle for evaluation of Hubbard interactions, overlaps and various matrix elements in a simple and efficient manner. 
\end{abstract}

\maketitle

\section{Introduction}
\label{sec:introduction}
Bloch theory in its original form described particle states in periodic potentials using a (quasi)momentum picture. It became clear later that it is beneficial to develop a position-space picture. The latter approach, introduced by Wannier~\cite{PhysRev.52.191}, uses localized functions which reflect the spatial structure of the potential and serve as ``hulls'' of the Bloch states. Owing to its conceptual simplicity, the Wannier function proved to be an indispensable tool for the study of a variety of research areas. One recent highlight is the role played by Wannier functions in theories of polarization in crystalline solids~\cite{PhysRevB.47.1651}. An extensive overview of these and other developments can be found in Ref.~\cite{RevModPhys.84.1419}. 

This work is motivated by recent efforts to use Wannier functions for understanding the states of cold atoms in optical lattices~\cite{PhysRevLett.81.3108}. Similar to previous work in regular solids, where Wannier functions have been used for first-principles calculations of parameters of the Hubbard model~\cite{PhysRevB.65.075103}, the parameters of the low-energy Hubbard-type Hamiltonian of cold bosonic atoms in optical lattices can be efficiently evaluated in Wannier basis. In the application to cold atoms, optical lattices often result in separable periodic potentials. In this case, instead of treating the full three-dimensional problem, one may consider a set of one-dimensional potentials. This makes developing a good understanding of Wannier functions for one-dimensional potentials particularly important.

\begin{figure}[h!tbp]
  \centering
  \begin{tikzpicture}
    \node[inner sep=0, outer sep=0] (plot) at (0,0) {\includegraphics[width=0.95\columnwidth]{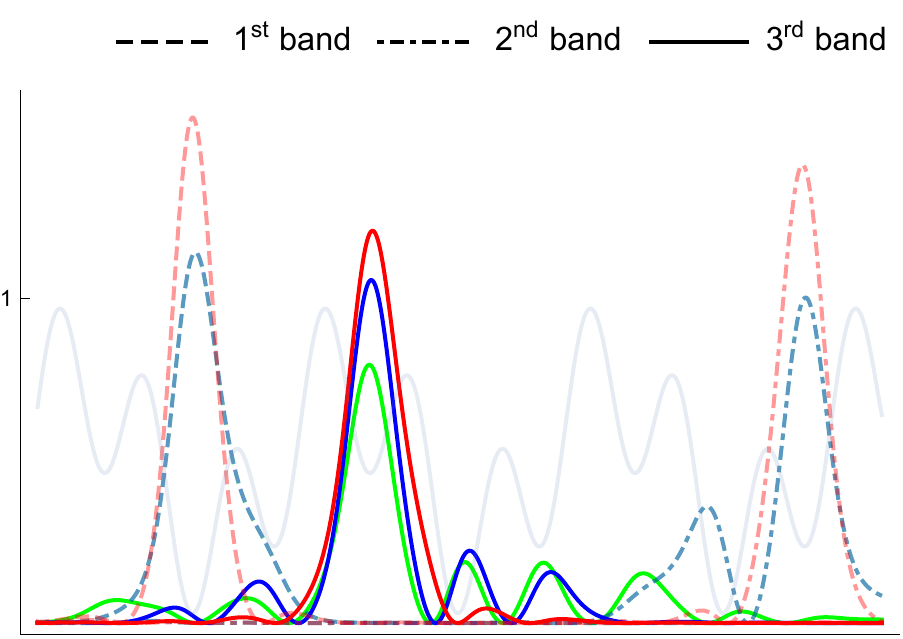}};
    \draw (plot.north west) ++(10pt,-10pt) node  {$|\psi(x)|^2$};
    \node[below] at (plot.south) {$x$};
  \end{tikzpicture}
  \caption{Maximally localized multi-band Wannier functions (red curves) for the lowest three bands in a triple well potential, similar to that proposed in Ref.~\cite{PhysRevLett.111.225301}. The functions are eigenstates of the $\Pi x \Pi$ matrix, as detailed in Section~\ref{sec-2-4}. The maximally localized \emph{single-band} wavefunctions (blue curves) are used to find the matrix elements of $\Pi x \Pi$ from the expressions in Section~\ref{sec-2-2}. Naive single-band Wannier functions with arbitrary phase on constituent Bloch waves (green curves) are used to find maximally localized single-band wavefunctions (see Section~\ref{sec-2-3}). Note the states in the highest band benefit the most from the localization procedure. The form of the potential is shown for reference.}
  \label{fig:process-figure}
\end{figure}

In the case of a single isolated band, for Bloch state in band $n$ and with quasi-momentum $\vec{k}$ $\ket{b_{n\vec{k}}}$, the general Wannier function may be written
\begin{equation}
  \label{eq:gen-single-band-wannier}
  \ket{\vec{s};n} \propto \int_{BZ} d\vec{k} e^{- i \vec{k} \cdot \vec{s}} \ket{b_{n\vec{k}}},
\end{equation}
where $\vec{s}$ is a vector to a lattice point and $\int_{BZ}$ indicates an integral over the first Brillouin zone. Already the strong non-uniqueness of the wavefunctions appears; each state $\ket{b_{n\vec{k}}}$ may be independently assigned an arbitrary $\vec{k}$ and $n$ dependent phase $\phi_n(\vec{k})$. If multiple bands need to be treated this indeterminacy only increases, for now there is also a free unitary transformation that can be done within the subspace of the relevant bands at each $\vec{k}$. Note that the need to treat multiple bands in one dimension, even in the absence of degeneracy, is not purely academic; for example, as long as the bands are relatively close, obtaining physically relevant Bose-Hubbard parameters for cold atoms requires considering all bands a particle may occupy. More specifically, a single-well potential as treated in detail in~\cite{PhysRevLett.81.3108} only requires single-band Wannier functions and does not benefit strongly from this analysis, but a recent proposal in Ref.~\cite{PhysRevLett.111.225301} to use a triple-well potential to create a quantum spin Hall Hamiltonian offers an example that requires maximally localized Wannier functions across the three lowest bands to obtain accurate Bose-Hubbard parameters. This proposal motivates the form of the sample potential in Figure~\ref{fig:process-figure}.

Methods of varying degree of complexity and generality exist for treating these functions. At one end of the spectrum, Kohn proves general analytic properties of Wannier functions in a single band~\cite{PhysRev.115.809}; of particular interest, he demonstrates the existence of (by appropriate choice of phase) exponentially localized Wannier functions. The supplementary material of Ref.~\cite{PhysRevLett.111.185307} details a construction from diagonalization of a projection operator similar to the one introduced in Section~\ref{sec-2-1}, but only treats the case of a single 1D band in general. A particular example with two bands, using the same projection operator as in this paper, is also given, but the problem is not treated for arbitrary problems; furthermore, maximally localized 1D Wannier states are not used as an intermediate basis for the projection operator. Finally, Marzari et al. present a general method for constructing the maximally localized Wannier functions in any dimensionality and band structure~\cite{PhysRevB.56.12847} that uses a minimization scheme.

In this paper we detail a constructive scheme for maximally localized Wannier functions in one-dimensional potentials; first an integral formula for single-band wavefunctions, then a diagonalization scheme for the general problem. Although our method is strictly less general than that in \cite{PhysRevB.56.12847}, it is computationally simpler and in particular does not rely on a minimization scheme, so for given precision is guaranteed to converge on a unique answer regardless of implementation and problem conditions. In fact, the precision of this method is straightforward to adjust, simply by increasing a number of matrix elements calculated (see Section~\ref{sec-2-4}). Furthermore, the only required non-trivial operation is matrix diagonalization, although integration may be used to improve the results on practical grounds; both of these operations are well-studied and have many efficient implementations. We also use (optionally maximally) localized single-band Wannier states (for which we provide a simple numerical construction) as our intermediate basis; this affords practical advantages in accuracy and convergence (see Section~\ref{sec-3}), conceptually abstracts the localization problem from the one of finding an initial basis for a particular Hamiltonian, and carries the formal advantage that the expectation of position is well-defined. Effectively, our method shares the scaling, uniqueness, simplicity, and convergence advantages cited by Ref.~\cite{PhysRevLett.111.185307}, while remaining sufficiently general to completely treat 1D periodic potentials of arbitrary complexity to high accuracy.

For the remainder of the paper we will restrict ourselves to one-dimensional potentials.

\section{Generalized Wannier functions}
\label{sec-2}

First, we introduce some essential concepts and notations. We assume Hamiltonian $H$ having periodicity $\pi / k_b$. Bloch's theorem then lends eigenstates $\ket{b_{nk}} = Q_k(x) \ket{u_{nk}}$ such that $Q_k(x) = e^{ikx}$ and $\braket{x}{u_{nk}} = \braket{x + \pi / k_b}{u_{nk}}$. We take $\braket{x}{u_{nk}}$ normalized within a unit cell, so that the corresponding Bloch wave normalization is $\braket{b_{n'k'}}{b_{nk}} = 2 k_b \delta (k - k') \delta_{nn'}$. 

In practice, if we further take the form of the Hamiltonian to be $H = p^2 / 2m + V(x)$, from $[p,Q_k(x)] = \hbar k Q_k(x)$ we have an eigenvalue equation $H' \ket{u_{nk}} = E_n \ket{u_{nk}}$ where
\begin{equation}
\label{eq:unk-h}
H' = \frac{(\hbar k + p)^2}{2m} + V(x).
\end{equation}
By expanding $H'$ in the basis $\braket{x}{q_m} = \sqrt{k_b/\pi} \exp{i2 k_b m x}, m \in \mathbb{Z}$ (complete for all $\ket{u_{nk}}$ by virtue of their periodicity) we may find energy structure and expressions for $\braket{x}{u_{nk}}$ to the desired precision (by choosing a large enough range for $m$). Note though that this is not necessary for the argument that follows.

In general, we want to consider some finite subset $B$ of the energy bands. To that end, we consider an operator $M(k)$ such that $\ket{b_{nk}'} = M(k) \ket{b_{nk}}$ where the matrix elements $\bra{b_{n'k''}} M(k) \ket{b_{nk'}}$ are non-zero only if $k' = k'' = k$ and $n,n' \in B$. We take $M(k)$ to be unitary in the subspace of $\ket{b_{nk'}}$, $n \in B$, $k' = k$. Then the $\ket{b_{nk}'}$ have the same normalization and orthogonality properties, and translation eigenvalues as the $\ket{b_{nk}}$. They are not generally energy eigenstates, but may still be written as $\braket{x}{b_{nk}'} = Q_k(x) \ket{u_{nk}'}$ where $\ket{u_{nk}'} = M(k) \ket{u_{nk}}$.

We define generalized Wannier functions as
\begin{equation}
\label{eq:gwdef}
\ket{s;n} = \frac{1}{2k_b} \int_{BZ} dk e^{-i s \pi k / k_b} M(k) \ket{b_{nk}}
\end{equation}
with the normalization condition $\braket{s';n'}{s;n} = \delta_{s's} \delta_{n'n}$. A useful operator will be the projection onto the bands $B$:
\begin{equation}
\label{eq:projection}
\Pi = \sum_s \sum_{n \in B} \ket{s;n} \bra{s;n} = \frac{1}{2k_b} \sum_{n \in B} \int_{BZ} dk \ket{b_{nk}} \bra{b_{nk}}
\end{equation}
where the last equality shows the invariance of $\Pi$ with respect to choice of $M(k)$. We will also use $R = 1 - \Pi$

\subsection{Wannier spread}
\label{sec-2-1}

We find an invariant portion of the spread of the Wannier function, first done in \cite{PhysRevB.56.12847}, and use this to find a matrix whose eigenstates are precisely the maximally localized versions of (\ref{eq:gwdef}). We define the spatial spread in the usual way, summing over the bands of interest,
\begin{align}
\label{eq:spread}
  &\sum_{n \in B} \bra{0;n} (x - \langle x \rangle)^2 \ket{0;n} \\
  &= \sum_{n \in B} \bra{0;n} (x^2 - x \ket{0;n} \bra{0;n} x) \ket{0;n} \\
  &= \sum_{n \in B} \bra{0;n} x \left(1 - \Pi + \sum_{\substack{s ; m \in B \\ s,m \ne 0,n}} \ket{s;m} \bra{s;m}\right) x \ket{0;n} \\
  \label{eq:spread-f}
  &= \sum_n \bra{0;n} \Pi x R x \Pi \ket{0;n} + \sum_{n \in B} \sum_{\substack{s ; m \in B \\ s,m \ne 0,n}} | \bra{0;n} x \ket{s;m} |^2
\end{align}
where we arbitrarily, but without loss of generality, choose $s = 0$. Note that (where $\mathscr{T}(y)$ represents translation in $x$ by $y$) if $[A,\mathscr{T}(s \pi / k_b)] = 0 \forall s \in \mathbb{Z}$ then
\begin{align*}
  \trace(A) &= \sum_s \sum_n \bra{s;n} A \ket{s;n} \\
  &= \sum_s \sum_n \bra{0;n} \mathscr{T}^{\dagger} (s \pi / k_b) A \mathscr{T}(s \pi / k_b) \ket{0;n} \\
  &= \sum_s \sum_n \bra{0;n} A \ket{0;n}, 
\end{align*}
so in particular $\sum_n \bra{0;n} A \ket{0;n}$ is basis-invariant, a result first expressed in \cite{PhysRevB.56.12847}. From $[\Pi, \mathscr{T}(s \pi / k_b)] = 0$ and $[x,\mathscr{T}(s \pi / k_b)] = s \pi / k_b \mathscr{T} (s \pi / k_b)$, we then have that the first term in (\ref{eq:spread-f}) is basis-invariant. It is also clear that both terms are positive-definite, so minimizing the spread amounts to minimizing the (magnitude of the) second term:
\begin{equation}
\label{eq:variant-spread}
S = \sum_{n \in B} \sum_{\substack{s ; m \in B \\ s,m \ne 0,n}} | \bra{0;n} x \ket{s;m} |^2.
\end{equation}
We may also write the summand as $\bra{0;n} \Pi x \Pi \ket{s;m}$. This notation makes clear the fact that if our Wannier functions are eigenstates of the operator $\Pi x \Pi$, then (by the orthonormality conditions in Section \ref{sec-2}) this positive-definite sum will be zero, hence minimized.

\subsection{Matrix elements of $x$}
\label{sec-2-2}

In order to construct the matrix $\Pi x \Pi$, we require an expression for matrix elements of $x$ in the Wannier basis. We note that
\begin{multline}
\label{eq:x-to-deriv}
x e^{-i \pi s k / k_b} \braket{x}{b_{nk}'} = \\
-i \left ( \partial_k ( e^{- i \pi s k / k_b} \braket{x}{b_{nk}'}) -
e^{i k (x - \pi s / k_b)} \partial_k (\braket{x}{u_{nk}'}) \right ) \\ 
+ \frac{\pi s}{k_b} e^{- i \pi s k / k_b} \braket{x}{b_{nk}'}
\end{multline}
Substituting (\ref{eq:x-to-deriv}) into the expression for $\bra{s';n'} x \ket{s;n}$, we have
\begin{multline}
\label{eq:x-expectation}
\bra{s';n'} x \ket{s;n} = \\ \frac{1}{2k_b} \int_{BZ} dk e^{-i \pi (s - s') k / k_b} C_{n'n}(k) + \frac{\pi s}{k_b} \delta_{ss'}\delta_{nn'}
\end{multline}
where
\begin{equation}
\label{eq:cnn-def}
C_{n'n}(k) = i \int_U dx \braket{u_{n'k}'}{x} (\partial_k \braket{x}{u_{nk}})
\end{equation}
where $\int_U dx$ indicates an integral over the unit cell. To evaluate the spread (\ref{eq:variant-spread}) we split it into off-site intraband and band-mixing pieces:
\begin{align}
\label{eq:spread-parts}
S_1 &= \sum_{n \in B} \sum_{s \ne 0} | \bra{0;n} x \ket{s;n} |^2 \\
S_2 &= \sum_{n \in B} \sum_s \sum_{m \ne n} | \bra{0;n} x \ket{s;m} |^2.
\end{align}
Noting that $C_{nn}(k)$ is real, we have
\begin{align}
\label{eq:explicit-s1}
\sum_{s \ne 0} |\bra{0;n} &x \ket{s;n}|^2 \\
&= \frac{1}{2k_b} \int_{BZ} dk C_{nn}(k)^2 - \left ( \frac{1}{2k_b} \int_{BZ} dk C_{nn}(k) \right )^2 \\
&= \frac{1}{2k_b} \int_{BZ} dk \left ( C_{nn}(k) - \frac{1}{2k_b} \int_{BZ} dk C_{nn}(k) \right )^2
\end{align}
and
\begin{equation*}
\sum_s | \bra{0;n} x \ket{s;m} |^2 = \frac{1}{2k_b} \int_{BZ} dk |C_{nm}(k)|^2.
\end{equation*}

\subsection{Maximally localized single-band wave-functions}
\label{sec-2-3}

Suppose our unitary matrix $M(k)$ does no band-mixing, i.e. $\bra{b_{n'k}} M(k) \ket{b_{nk}}$ is non-zero only for $n' = n$. In this case $M(k)$ consists of pure phases indexed by energy on the diagonal, $e^{i \phi_n(k)}$. If $C_{n'n}(k)$ are defined as in (\ref{eq:cnn-def}) with $M(k) = \matid$, the new $C'_{n'n}(k)$ matrix with $M(k)$ being ``diagonal'' are
\begin{equation*}
C'_{n'n}(k) = - \partial_k \phi_n (k) \delta_{n'n} + e^{i(\phi_n(k) - \phi_{n'}(k))} C_{n'n}(k).
\end{equation*}
Since $S_2$ only depends on $|C_{n'n}(k)|^2$, $n' \ne n$, it is not affected by such a mixing matrix $M(k)$. $S_1$, on the other hand, is non-trivially affected. By the argument in Section \ref{sec-2} $S_1 = 0$ for the most localized wave-function, so (\ref{eq:explicit-s1}) offers the condition $C_{nn}(k) = (1/2k_b) \int_{BZ} dk C_{nn}(k)$, from which we can extract the ideal phase factors $\phi_n(k)$ (up to an irrelevant global phase, fixing $k_0$ and choosing $\phi_n(k_0) = 0$):
\begin{equation*}
\phi_n(k) = \int_{k_0}^k dk C_{nn}(k) - \frac{k - k_0}{2k_b} \int_{k_0}^{k_0 + 2k_b} C_{nn}(k).
\end{equation*}
Applying these phases to each Bloch wave at the particular energy $n$ and quasi-momentum $k$ gives maximally-localized single-band wave-functions, for an example see Figure~\ref{fig:process-figure}.

\subsection{Maximally localized Wannier functions}
\label{sec-2-4}

Having optionally preformed the procedure described in Section \ref{sec-2-3}, we may find matrix-elements of $\Pi x \Pi$ in a maximally-localized single-band basis using (\ref{eq:x-expectation}) and the following explicit expressions. Although the single-band minimization is not necessary to find matrix elements of $\Pi x \Pi$, we intuitively expect \ylnotes{Provable, can add argument or possible citation} that working in a maximally localized single-band basis will in turn reduce the mixing of these bases in the maximally localized multi-band basis functions. This helps to decouple the accuracy of the construction from the dimensionality used for $\Pi x \Pi$; only elements of the basis in which $\Pi x \Pi$ is expressed that have their mean position close to the site that corresponds to a given eigenvector will have non-zero coefficients as given by the numerical representation of that eigenvector. This in turn means that past some dimensionality for $\Pi x \Pi$, increasing the number of matrix elements calculated will not significantly change the states that contribute to the maximally localized functions. In any case, once we have built the matrix $\Pi x \Pi$, we can use any suitable method to find eigenvectors in the basis $\Pi x \Pi$ is constructed; these eigenvectors will be the maximally localized Wannier functions.

Lastly, as an aside, we note an interesting question that stems from the above analysis. If we choose not to limit ourselves to diagonal operators as in Section \ref{sec-2-3}, we find that for a given $M(k)$ and old $C_{n'n}(k)$, the new $C'_{n'n}(k)$ satisfy
\begin{equation}
\label{eq:cn-matrix}
C'(k)^T = i (\partial_k M(k)) M(k)^{\dagger} + M(k) C(k)^T M(k)^{\dagger}.
\end{equation}
In the maximally localized case, $C'(k)$ is diagonal, with a given diagonal element constant over $k$ (see the expressions in \ref{sec-2-2}). In fact, it is just necessary to find $M(k)$ such that $C'(k)$ is diagonal, as the diagonal elements can then be adjusted to optimal values as in Section \ref{sec-2-3}. The author does not know of a general way to solve for $M(k)$ in this expression, but such a method could provide a possibly faster numerical alternative to diagonalizing the $\Pi x \Pi$ matrix.

\section{Conclusions}
\label{sec-3}

We have presented an algorithm for constructing maximally localized one dimensional Wannier functions. Although not the most general procedure, this prescription has advantages of computational simplicity, uniqueness of outputs with respect to implementations and inputs, and straightforward adjustment of precision by increasing the number of matrix elements of the quantity $\Pi x \Pi$ computed before diagonalization. 
The above construction of Wannier functions will have a number of applications to cold atoms in separable optical lattice, since well-localized functions make evaluation of Hubbard interactions, overlaps and various matrix elements simple and efficient. Beyond their utility for optical lattices, this method can provide new insights and better understanding in other one dimensional problems.

\section{Acknowledgments}
\label{sec:acknowledgements}

We acknowledge the support of Professor Wolfgang Ketterle, and thank William Cody Burton for useful discussion. Finally, we thank Professor Leonid Levitov for invaluable guidance and advice.

\bibliography{1dmlwf}

\begin{thebibliography}{1}

\bibitem{PhysRev.52.191}
Gregory~H. Wannier.
\newblock The structure of electronic excitation levels in insulating crystals.
\newblock {\em Phys. Rev.}, 52:191--197, Aug 1937.

\bibitem{PhysRevB.47.1651}
R.~D. King-Smith and David Vanderbilt.
\newblock Theory of polarization of crystalline solids.
\newblock {\em Phys. Rev. B}, 47:1651--1654, Jan 1993.

\bibitem{RevModPhys.84.1419}
Nicola Marzari, Arash~A. Mostofi, Jonathan~R. Yates, Ivo Souza, and David
  Vanderbilt.
\newblock Maximally localized wannier functions: Theory and applications.
\newblock {\em Rev. Mod. Phys.}, 84:1419--1475, Oct 2012.

\bibitem{PhysRevLett.81.3108}
D.~Jaksch, C.~Bruder, J.~I. Cirac, C.~W. Gardiner, and P.~Zoller.
\newblock Cold bosonic atoms in optical lattices.
\newblock {\em Phys. Rev. Lett.}, 81:3108--3111, Oct 1998.

\bibitem{PhysRevB.65.075103}
I.~Schnell, G.~Czycholl, and R.~C. Albers.
\newblock Hubbard-$u$ calculations for cu from first-principle wannier
  functions.
\newblock {\em Phys. Rev. B}, 65:075103, Jan 2002.

\bibitem{PhysRevLett.111.225301}
Colin Kennedy, Georgios Siviloglou, Hirokazu Miyake, William Burton, and
  Wolfgang Ketterle.
\newblock Spin-orbit coupling and quantum spin hall effect for neutral atoms
  without spin flips.
\newblock {\em Phys. Rev. Lett.}, 111:225301, Nov 2013.

\bibitem{PhysRev.115.809}
W.~Kohn.
\newblock Analytic properties of bloch waves and wannier functions.
\newblock {\em Phys. Rev.}, 115:809--821, Aug 1959.

\bibitem{PhysRevLett.111.185307}
Thomas Uehlinger, Gregor Jotzu, Michael Messer, Daniel Greif, Walter
  Hofstetter, Ulf Bissbort, and Tilman Esslinger.
\newblock Artificial graphene with tunable interactions.
\newblock {\em Phys. Rev. Lett.}, 111:185307, Oct 2013.

\bibitem{PhysRevB.56.12847}
Nicola Marzari and David Vanderbilt.
\newblock Maximally localized generalized wannier functions for composite
  energy bands.
\newblock {\em Phys. Rev. B}, 56:12847--12865, Nov 1997.

\end{thebibliography}
\bibliographystyle{unsrt}

\end{document}